# Securing U-Healthcare Sensor Networks using Public Key Based Scheme[*]

Md. Mokammel Haque, Al-Sakib Khan Pathan, and Choong Seon Hong
Department of Computer Engineering, Kyung Hee University, Korea 449-701
{malinhaque, spathan}@networking.khu.ac.kr and cshong@khu.ac.kr

*Abstract* — Recent emergence of electronic culture uplifts healthcare facilities to a new era with the aid of wireless sensor network (WSN) technology. Due to the sensitiveness of medical data, austere privacy and security are inevitable for all parts of healthcare systems. However, the constantly evolving nature and constrained resources of sensors in WSN inflict unavailability of a lucid line of defense to ensure perfect security. In order to provide holistic security, protections must be incorporated in every component of healthcare sensor networks. This paper proposes an efficient security scheme for healthcare applications of WSN which uses the notion of public key cryptosystem. Our entire security scheme comprises basically of two parts; a key handshaking scheme based on simple linear operations and the derivation of decryption key by a receiver node for a particular sender in the network. Our architecture allows both base station to node or node to base station secure communications, and node-to-node secure communications. We consider both the issues of stringent security and network performance to propose our security scheme.

## 1. Introduction

Pervasive or Ubiquitous Healthcare (U-Healthcare) is viewed as a wide-scale deployment of healthcare facilities among patients, physicians, and other health-care workers to transport medical information accurately and securely anywhere, at anytime. It integrates unique capabilities of emerging wireless sensor network (WSN) technologies that can support a wide range of applications including telemedicine, patient monitoring, location-based medical services, emergency response, personalized monitoring, and incentive management [1]. But privacy and security requirements for these applications are very diverse as they are based on different usage scenarios ranging from pre-hospital, in-hospital, ambulatory, home monitoring, to database collection for long-term trend analysis [2]. Also the resource constriction and constant topology evolving characteristics of WSN further aggravate the security challenges. Hence, a properly designed end-to-end security architecture is the primary necessity for efficient utilization of WSN in U-Healthcare.

Ensuring complete and a good level of security for such types of networks however, is not a trivial task. As these types of networks use wireless communications, the threats and attacks against them are more diverse and often very large in scale. It is practically impossible to deal with all sorts of security threats with a single mechanism. Instead, a combination of different security schemes for a single network could be the solution [3].

Considering the special features of wireless sensor networks, in this paper, we propose an efficient public key based security scheme for WSN. In our scheme, we use pseudoinverse matrix for the first part while the second part is a simple method for transferring decryption key to the receiver node. Our analysis and simulation results show that our scheme demonstrates a considerable gain in the level of security and is suitable enough to be employed with the current generation sensor nodes.

The rest of the paper is organized as follows: Section 2 states the related works, Section 3 presents our proposal, Section 4 states the performance analysis, and Section 5 concludes the paper with future research directions.

## 2. Related Works

So far a few security schemes have been proposed for healthcare sensor networks. CodeBlue [4] implements ECC on Mica2 mote using only integer arithmetic. CodeBlue comprises a suite of protocols for emergency response. WBAN (Wireless Body Area Network) infrastructure presented in [5] is to develop a wearable health status monitoring system.

In [6], [7], and [8], PKI (Public Key Infrastructure) based security solutions are proposed for healthcare. [6] emphasizes on multi-layered security infrastructure that copes with the strong user authentication procedures based on smart cards, digital certificates, and PKI systems. The system proposed in [7] comprises of six subsystems. In the proposed architecture, the authors used radio frequency (RFID) tag, wearable electrocardiogram (ECG) sensor, smart card, grid computing, PhysioNet, wired/wireless network, and public-key infrastructure technologies for Ubiquitous RFID Healthcare System for WSN. [8] proposes wireless PKI security scheme for 3G mobile communication. The scheme consists of a trusted server and a dual public key cryptosystem to provide

[*] This work was supported by the Korea Research Foundation Grant funded by the Korean Government (MOEHRD) (KRF-2006-521-D00394). Dr. CS Hong is corresponding author.

end-to-end security between mobile clients and the Healthcare Information System (HIS).

Typically the appropriateness of asymmetric systems for authenticity, integrity protection, and non-repudiation is widely accepted; alike for symmetric systems in data confidentiality. In this paper, we propose an asymmetric approach for confidentiality protection to demonstrate the merit of PKI and the setting for applying the scheme is a healthcare sensor network where security of medical data is one of the most important concerns.

## 3. Our Security Scheme

### 3.1 Network Model

We consider a hierarchical model of the network where the primary components are; Patient (PT), Healthcare Service System (HSS), and Secure Base Station (SBS). HSS could be doctor, nurse or any other medical staffs. There are multiple SBSs installed in the network within the hospital compound. The SBSs could be replaced if necessary. They are connected with each other using wired or wireless technology and have unlimited energy sources. They can communicate among themselves if necessary. Any HSS or a PT is considered as a node in the model. The sensors which carry medical data are either wearable or attached with the HSSs or PTs.

### 3.2 Pseudoinverse Matrix

The pseudoinverse matrix or generalized inverse matrix [9] has a very nice property that could be used for cryptographic operations. It is well known that, a nonsingular matrix over any field has a unique inverse. For a general matrix of dimension $k \times n$, there might exist more than one generalized inverse. This is denoted by, $M(k,n) = \{A: A \text{ is a } k \times n \text{ matrix}\}$. Let, $A \in M(k,n)$. If there exists a matrix $B \in M(n,k)$ such that, $ABA = A$ and $BAB = B$, then each of $A$ and $B$ is called a generalized inverse matrix (or pseudoinverse matrix) of the other. In this paper, we use the notation $A_g$ to denote the generalized inverse matrix or pseudoinverse matrix of $A$. We use pseudoinverse matrix for deriving the pairwise secret key between any HSS and SBS or between any PT and SBS.

### 3.3 Bilateral Key Handshaking between SBS and PT (or HSS)

Let a PT (or HSS) wants to join the network. For secure communications with SBS, it needs to derive a pairwise secret key with SBS(s). For this, following operations are performed:

1. The sensor ($s_i$) associated with PT (or HSS) randomly generates a matrix $X$ (dimension $m \times n$) and its psuedoinverse matrix $X_g$. These matrices are kept secret in the node.

2. Then $s_i$ calculates $X_g X$ and sends it to the nearest SBS.

3. In turn, SBS randomly generates another matrix $Y$ with dimension $n \times k$, and finds out its pseudoinverse matrix $Y_g$. These matrices are also kept secret in the SBS.

4. SBS calculates $X_g XY$ and $X_g XYY_g$. Then it sends the resultant matrices to $s_i$.

5. Upon receiving the products of matrices from SBS, $s_i$ calculates, $XX_g XYY_g = XYY_g$ and sends it back to SBS.

6. Now, both the node $s_i$ and the SBS can compute the common secret key. $s_i$ gets it by calculating $X(X_g XY) = XY$ and the SBS gets it by calculating $(XYY_g)Y = XY$. Both of these outcomes (XY) are the same matrix with dimension $m \times k$. Figure 1 shows the scheme and network structure.

The pairwise shared secret key (XY) with the SBS can also be stored in its memory before its deployment and can be used as the encryption key for the encryption-decryption phase.

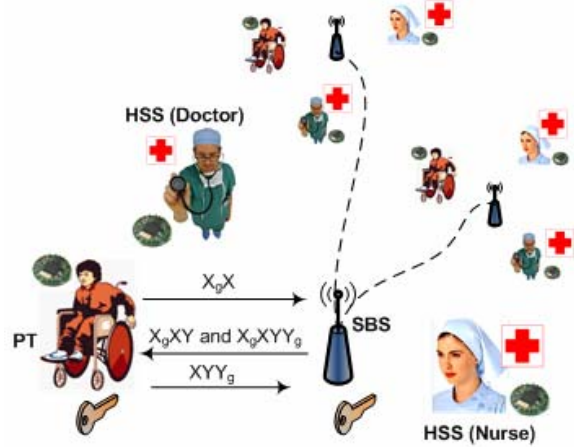

**Figure 1. Pairwise Secret Key Derivation**

### 3.4 Encryption and Decryption of Data for PT-to-HSS or HSS-to-PT Communications

The main module in secure node-to-node communication is a central key generator (CKG) which is located at the SBS. The CKG helps any node in the network to decrypt the received encrypted messages from other nodes. If a node $n_A$ wants to send message securely to another node $n_B$, it uses its pairwise key with the SBS for encrypting the message. Say for example, the encrypted message sent from $n_A$ to $n_B$ is $E_{XY}(M)$. Here, $M$ is the message sent from the sender to the receiver. $E_{XY}$ means that the message is encrypted with the key $XY$, which is actually the shared secret key between the SBS and the sender (either a PT or a HSS). Upon receiving the encrypted message, $n_B$ places its own identity and the identity of the sender to the CKG. In turn, CKG generates a decryption key and transmits it to $n_B$ encrypting it with the secret shared key that it has with $n_B$. As the CKG in the SBS has prior knowledge about the shared secret keys of both the nodes, it uses that knowledge to generate the decryption key. Now, $n_B$ first decrypts the encrypted message (the message sent from SBS) with its shared key with SBS, finds out the decryption key, and uses that key to decrypt the message sent from node $n_A$.

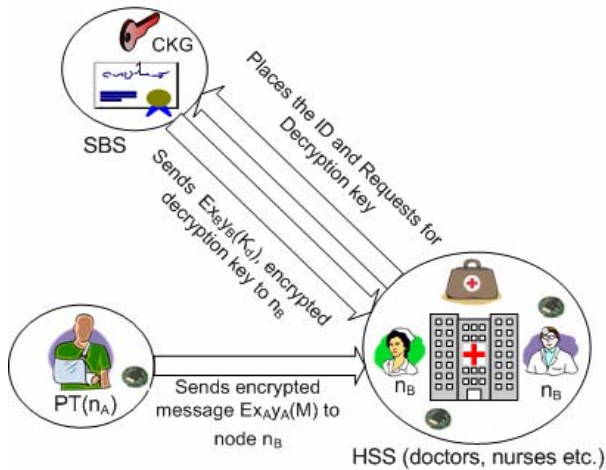

**Figure 2. Encryption and Decryption of Message by Two Communicating Nodes**

Figure 2 shows the secure communication method between two nodes in the network. In the figure, $X_A Y_A$ is the shared secret key between $n_A$ and SBS, $X_B Y_B$ is the shared secret key between $n_B$ and the SBS, and $K_d$ is the decryption key provided by the CKG to the receiving node.

# 4. Performance Evaluation

## 4.1 Initial Analysis

In our scheme, any node $s_i$ sends the *SBS* an $n \times n$ matrix which is of $n^2$ bits. In turn, the *SBS* sends an $n \times k$ matrix and an $n \times n$ matrix. For this the total number of bits passed for the matrices is, $n^2 + nk = n(n+k)$ bits. Again, the node $s_i$ sends the *SBS* an $m \times n$ bits. So, total number of bits for the matrices transmitted for deriving the shared key in the whole key handshaking process is,

$$n^2 + n(n+k) + mn$$
$$= n(n + n + k + m)$$
$$= n(2n + k + m)$$

All the calculations here are linear and can be performed very easily. Moreover, our scheme is adequately secure as capturing the messages like $X_g X$, $X_g XY$, $X_g XYY_g$, and $XYY_g$ could not be in any way helpful to construct the locally computed secret shared key $XY$.

## 4.2 Simulation

We have analyzed our PKC-based scheme in terms of energy cost, memory cost, security and scalability. In our simulation, we considered the specifications of Berkeley/Crossbow MICA2dot [10] motes. These motes are equipped with 8-bit ATmega128L microcontrollers with 4 MHz clock speed, 128 KB program memory and Chipcon CC1000 low-power wireless transceiver with 433-916 MHz frequency band. The major power consumers in this mote are the processor and the wireless transceiver. During the transmission and reception operations, the microcontroller is turned on alongside the wireless transceiver. According to our calculations, the cost of transmission of one byte is 59.2 μJ while the reception operation takes about half of the transmission cost (28.6 μJ). The power to transmit 1 bit is equivalent to roughly 2090 clock cycles of execution of the microcontroller. In our case, we considered a packet size of 41 bytes (payload of 32 bytes, header 9 bytes). With an 8 byte preamble (source and destination address, packet length, packet ID, CRC and a control byte) for each packet we found that, to transmit one packet $49 \times 59.2 = 2.9008$ mJ ≈ 2.9 mJ energy is required. Accordingly, the energy cost for receiving the same packet is $49 \times 28.6 = 1.4014$ mJ ≈ 1.4 mJ. Considering the same packet size for all the network operations, to set up a shared secret key with the base station each node needs (two transmissions and one reception) $((2 \times 2.9) + 1.4) = 7.2$ mJ of energy. This cost is one time cost as once the shared secret key is derived, it could be used for the entire lifetime of the network unless the key is exposed or the node quits the network.

For node to node communication, the sender needs one transmission (2.9 mJ) and the receiver needs two receptions and one transmission $(((2 \times 1.4) + 2.9) = 5.7$ mJ). As a whole, the entire scheme could be well-afforded by the energy resources of the current generation sensor nodes.

## 4.3 Comparison

We compare our PKC-based scheme with C4W [11] and the one proposed in [12], which use simplified version of SSL handshake. Considering the energy consumption for communications, our scheme stays in the middle of other two schemes (shown in table 1).

**Table 1. Communication cost for our scheme and other PKC-based schemes**

| PKC-based Schemes | Communication Cost | |
|---|---|---|
| | Sender ($n_A$) | Receiver ($n_B$) |
| C4W | 6.3 mJ | 4.8 mJ |
| Our Scheme | 10.1 mJ | 5.7 mJ |
| SSSL | 19.4 mJ | 19.6 mJ |

For supporting our scheme, for sender node ($n_A$), the numbers of transmissions and receptions are 3 and 1 respectively which take $((3 \times 2.9) + 1.4) = 10.1$ mJ) of energy in total. In case of receiver node ($n_B$), it is 5.7 mJ for 1 transmission and 2 receptions. Figure 3 shows a comparative graph in terms of communication cost among these schemes.

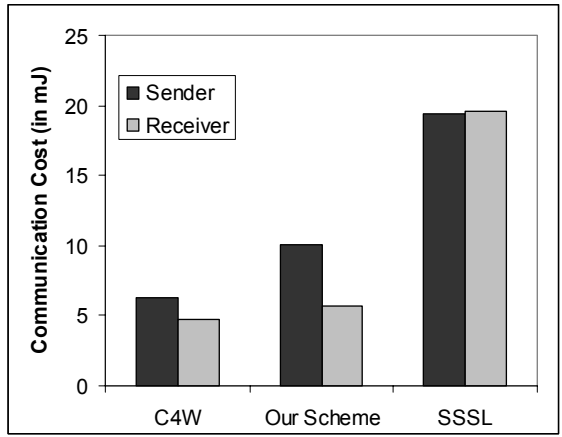

Figure 3. Communication cost for different PKC-based schemes

Though considering the communication cost, C4W exhibits better cost effectiveness than our scheme and SSSL, it requires pre-storing of all parameters before deployment. This in fact causes memory exploitation which is not present in our scheme. So, in terms of memory usage our PKC-based scheme is better and its communication cost is satisfactory.

**4.4 Further Comments**

In encryption decryption phase two messages are transmitted over public channel. When the receiver node needs the decryption key to decrypt a message from a particular sender node, it requests the SBS for the corresponding decryption key. In return, SBS encrypts the decryption key with the shared secret key of the receiver node. As the shared secret key is not known to any other node in the network, the decryption key for that particular sender-receiver pair could not be exposed. Now, the problem arises if the shared secret key of a node in the network is somehow compromised. In such a case, the base station revokes the shared key and the key handshaking process is re-initiated for that particular node. If such a compromise happens, even in that case, only one node is affected in the network while all other nodes could properly operate with confidential message transmission.

As any node can get a corresponding decryption key from the SBS (base station) for any sender node in the network, any pair of nodes in the network could communicate between themselves maintaining the high level of security. As mentioned earlier, for base station to node communications or node to base station communications, the shared secret key derived from the handshaking process is used which takes very little computation and message transmission.

## 5. Conclusion and Future Works

This paper proposes a security scheme for U-Healthcare Sensor Network (U-HSN) which uses the notion of public key cryptosystem. It should be mentioned that, our approach is well-scalable as any number of new sensors could be added to an existing wireless sensor network based on the requirements. Any newly added node could derive a shared secret key with the base station using the key handshaking process and then it could be used for node-to-node secure communications. In future, we would like to analyze our scheme in details considering different types of installation scenarios and security requirements of the application at hand.